\documentclass[aps,pre,showpacs,twocolumn]{revtex4}
\usepackage{amssymb}
\usepackage{graphicx}
\usepackage{colordvi}
\usepackage{color}

\begin{document}

\title{Dissipative periodic waves, solitons and breathers of the nonlinear Schr\"odinger equation with complex potentials}
\author{F. Kh. Abdullaev$^{1,2}$, V. V. Konotop$^{1,3}$, M. Salerno$^4$, and A. V. Yulin$^1$}
\affiliation{ $^1$Centro de F\'isica Te\'orica e Computacional, Faculdade de Ci\^encias,
Universidade de Lisboa,   Avenida Professor
Gama Pinto 2, Lisboa 1649-003, Portugal
\\
$^2$Physical - Technical Institute, Uzbek Academy of
Sciences, 2-b, G. Mavlyanov str., 100084, Tashkent, Uzbekistan
\\
$^3$Departamento de F\'isica, Faculdade de Ci\^encias,
Universidade de Lisboa, Campo Grande, Ed. C8, Piso 6, Lisboa
1749-016, Portugal
\\
$^4$Dipartimento di Fisica ``E.R. Caianiello'', CNISM  and INFN -
Gruppo Collegato di Salerno,
Universit\`a di Salerno, Via Ponte don Melillo, 84084 Fisciano (SA), Italy}

\date{\today}

\begin{abstract}
Exact solutions for the generalized nonlinear Schr\"odinger (NLS) equation with
inhomogeneous complex linear and nonlinear potentials  are
found. We have found localized and periodic solutions for a wide class
of localized and periodic modulations in the space of complex potentials and
nonlinearity coefficients. Examples of stable and unstable solutions are given. We also demonstrated numerically
the existence of stable
dissipative breathers in the presence of an additional
parabolic trap.
\end{abstract}

\maketitle
\input{epsf.tex} \epsfverbosetrue

\section{Introduction}
 Dissipative wave phenomena in nonlinear media with complex parameters  are attracting nowadays a great deal of attention.  They  appear naturally in optics of active media like, for example cavities, active fibers, etc.~\cite{optics} and in quantum mechanics  when   inelastic interactions of particles with external forces are accounted~\cite{QM}.  More recently, there  has been a particular interest in the dynamics of nonlinear waves  in periodic complex potentials due to possible applications to matter waves in absorbing optical lattices~\cite{Abd08,Bludov10} and to  periodic modulations of a complex refractive index in nonlinear optics~\cite{Musslim1,Staliunas}. In this last case  special attention was devoted  to the so called ${\cal PT}$ potentials  (i.e. complex potentials for which the nonlinear Schr\"odinger (NLS) equation becomes invariant under the parity and time-reversal symmetry)\cite{Bender},  whose remarkable linear properties are presently  intensively investigated~\cite{JPA,Longhi}. In particular, it has been  shown that in nonlinear media with
${\cal PT}$-symmetric {\it linear} damping and amplifications,
stationary localized and periodic states may exist \cite{Musslim2}. The stability problem of these nonlinear structures  is still an open problem.   Also, it is not investigated so far whether similar structures with real energies could also exist in a general complex potential, thanks to the  balance between nonlinearity, dispersion, gain and dissipation.

 In the present  paper we shall address these problems with a  twofold aim.
From one side we derive a set of exact soliton solutions of the  nonlinear Schr\"odinger (NLS)  equation with complex potentials and show that the nonlinearity management technique of the gain-loss profile can be an effective mechanism for providing stability.

 We find exact solutions by  adopting  an  inverse engineering approach, developed in~\cite{BK} for the case of the  conservative NLS equation with periodic coefficients. More specifically  we assume a specific pattern for the solutions and determine a posteriori the potentials which can sustain such  pattern as a solution.  Using this approach we determine a general class of potentials  which may support dissipative periodic waves and solitons in the form of elliptic functions. We show that some of these solutions my be stable with respect to small perturbations.

 From the other side,  we show that the derived solutions can be used to explore interesting dynamical behaviors of localized modes of the complex NLS equation. To this regard we investigate stable
breather solutions which originate from localized exact solutions  when a parabolic trap is switched on as additional real potential. Using the strength of the parabolic trap as a parameter we
show the occurrence of a bifurcation from an attractor center, corresponding to a stationary soliton, to a  limit cycle, corresponding to a stable breather mode.

The stability of stationary periodic solutions, single hump solitons,  and breathers   opens the possibility to experimentally observe such waves both in nonlinear optics and in BEC with optical lattices in presence of dissipative effects.

We remark that although the determination of the potentials from the
solutions is opposite to what usually occurs  in practical contexts
where  potentials are a priori fixed,  there are  physical situations in
which  an inverse approach can be experimentally implemented.
An example of this is the case of a Bose-Einstein condensates nonlinear optical lattices which is produced by means of spatial modulations of the interatomic scattering length. In the mean  field approximation  the system is described by the NLS equation with spatially modulated nonlinearity of the type considered in this paper. Since the modulation of the scattering length (nonlinearity)  in a real experiment is controlled by external magnetic fields via a Feshbach resonance, in principle one can construct arbitrary potentials to support specific prepared states, implementing thus an inverse engineering approach.
Another situation, where the potential can be adjusted in accordance with the profile designed a priori, is the nonlinear optics of waveguide arrays, where the thermo-optics effect is employed using heaters producing the temperature gradients properly distributed in space. This gives the hope that  some of the exact solutions  reported in this paper could be indeed observed in experimentally contexts.

 The paper is organized as follows. In section II we introduce the model equation and explain the inverse method used to determine the solutions. In Section III we apply the method to the case in which there are only linear real and complex  potentials. As an example we derive solutions in the potential which has the ${\cal PT}$-symmetry. We show that some of these solutions can be stable,   what is confirmed  by both the linear stability analysis and by numerical time evolutions of the complex NLS equation. In section IV we do similar studies for the case of nonlinear   lattices. In section V we show an example of exact stable solutions in correspondence of more general potentials, while in section  VI we use this solution to numerically show the existence of  dissipative breathers in the complex NLS equation with additional real parabolic trap. Finally, in section VII  the main results of the paper  are briefly summarized.

\section{The model and the method}
We consider the generalized NLS equation with
varying in space complex linear potential, $U_l(x)\equiv V_l(x)+iW_l(x)$, and complex nonlinearity
parameter $ U_{nl}\equiv V_{nl}(x)+iW_{nl}(x)$ (hereafter $V=$Re$U$ and $W=$Im$U$)
\begin{eqnarray}
\label{sys}
i\psi_t = -\frac{1}{2}\psi_{xx}+ \sigma |\psi|^2\psi +
U_l(x) \psi
+ U_{nl}(x) |\psi|^2\psi,
\end{eqnarray}
where $\sigma=\pm 1$ and is introduced in an explicit form in order to facilitate transition to the case of the homogeneous nonlinearity (just putting $V_{nl}(x),W_{nl}(x)\equiv 0$).
In the particular case of the ${\cal PT}$ symmetry the invariance imposes  restrictions on the  linear and complex  potentials:
namely $V_l,V_{nl}$ must be even and $W_l, W_{nl}$ must be odd  functions of the space variable, respectively.

We will be interested in the solutions of the form
$
\psi = A(x)\exp\left(i[\theta(x)-\omega t]\right),
$
where $A(x)$ and $\theta(x)$ are real amplitude and inhomogeneous phase of the mode.
Substituting this ansatz in  Eq.~(\ref{sys}) we
obtain the system of equations
\begin{eqnarray}
\label{sys2}
\omega A + \frac{1}{2}A_{xx}  - \frac{A}{2}v^2 - \sigma A^3 - V_l A - V_{nl}A^3 =0,\\
A_xv  +\frac 12  Av_{x}   -W_l A - W_{nl}A^3 = 0,
\label{sys3}
\end{eqnarray}
where $v\equiv\theta_x$.

Let us now assume that linear and nonlinear potentials $V_l(x)$ and $V_{nl}(x)$  are given, and pose the problem of designing dissipative terms  $W_l(x)$ and $W_{nl}(x)$ to obtain  a given solution $A(x)$~\cite{BK}. We emphasize that this choice is only for illustration proposes and any pair of the four functions $v_{l,nl}(x)$ and $W_{l,nl}(x)$ can be chosen as a priori given.

As first step, we prove that this is indeed possible, provided that $A(x)$ is a bounded function \footnote{To figure out whether our method could be extended for the case when $A(x)$ are singular, further investigations are required} satisfying the condition
\begin{eqnarray}
\label{cond1}
|f(x)|<\mbox{const},\quad\mbox{where}\quad f(x)\equiv A_{xx}/A
\end{eqnarray}
for all $x$.
Indeed, after dividing (\ref{sys2}) by $A$ we obtain and explicit expression for the velocity
\begin{eqnarray}
\label{auxil_1}
v^2(x)=  2(\omega -V_l) + f(x)-  2(\sigma  +  V_{nl}) A^2.
\end{eqnarray}
As it is clear this equation always has solutions for sufficiently large frequency $\omega$.
Next, substituting (\ref{auxil_1}) in (\ref{sys3}) we find the expression for the dissipative potentials
\begin{eqnarray}
\label{w}
W_l  +W_{nl}A^2 = \frac{1}{2A^2}\frac{d}{dx}(A^2 v)=\frac{A_xv}{A}  + \frac 12 v_{x}.
\end{eqnarray}
If we want to consider only nonsingular dissipative terms, then we have to impose one more constrain on the field $A$, which must be considered together with (\ref{cond1})
\begin{eqnarray}
\label{cond2}
\left| \frac{ A_{x} v}{A}\right|<\mbox{const},
\end{eqnarray}
and assume $v$ to be a smooth function of $x$.

In what follows we limit ourselves only by this kind of dissipation.
This allows us to indicate immediately several types of admissible solutions:
\begin{itemize}
\item{\em Type 1}: $A$ is bounded and has no zeros. These, for example are functions like dn$(x,k)$ etc.
\item {\em Type 2}: $A$ is bounded, has no zeros in any finite domain, but decays as $|x|\to \infty$. These are solutions like  $1/\cosh(x)$, $\exp(-x^{2})$, etc.
\end{itemize}
For these types of solutions condition (\ref{cond2}) is automatically satisfied if we restrict our consideration to bounded conservative potentials.

Further, we relax the requirements for $A$ allowing it to have zeros but we require that condition (\ref{cond2}) holds. Then the condition (\ref{cond1}) leads to another type of solutions:
\begin{itemize}
\item{\em Type 3}: $A$ has zeros which coincide with those  of $A_{xx}$. These are solutions of the type $\sim$sn$(x,k) $, etc.
\end{itemize}
All solutions reported below belong to one of these three types. We would like to notice here that $A$ must not necessarily be taken in the form of elliptic functions. We choose elliptical functions just as an example representing periodic solutions with a given parameter (the elliptic modulus). Alternatively we could choose the field $A$ in the form of, for instance, Gaussian function. Another remark we would like to make here is that if $A$ is a solution of the homogeneous NLS equation, then the coefficients $V_l$ and $V_{nl}$ are constants and one has to determine $W_{l}$  from either eq. (\ref{sys3}) or (\ref{w}) for a chosen $W_{nl}$. Notice, that this procedure implies some freedom, in the sense that one can also think of determining $W_{nl}$ for chosen $W_{l}$. However the latter might require some additional conditions on the linear dissipative potential $W_{l}$ because $sn$-like solutions have zeros. In general terms this case is mentioned in the paper as a solution of Type 3.

In closing this section we remark that from the above analysis it follows
that for all constructed potentials one can define a function $f(x)$ such that the nonlinear solutions  are also solutions of the {\em linear} problem
$A_{xx}-f(x)A=0$. This observation leads to an immediate interesting conclusion.
Assume that $ V_{nl}(x)<0$ and choose $f(x)=-2(\omega-V_l)$. Then, for a periodic  functions $V_l$, the solution $A$ is nothing but a {\em linear Bloch function} of the linear potential $2V_l(x)$. Indeed, in this case $A_{xx}-2V_l(x)A=-2\omega A$ and  $v(x)=\sqrt{2|V_{nl}|}A$, which means that condition (\ref{cond2}) is satisfied.

\section{ Case of linear complex potentials}

According the scheme described above one can design the dissipation of the system in a way to provide the any desired field configuration supported by the given linear and nonlinear conservative potentials. While the zoo of available nonlinear patterns is not limited, physically the most relevant cases are those corresponding to the stable solutions. Below we present several examples of the stable field distribution, completing the analysis with a contrasting behavior of unstable patterns.

Before going into detail we also notice that the diversity of the physical parameters at hand makes the analysis of particular cases cumbersome. It turns out, however that at least one of them can be scaled out. In particular, the amplitude of the nonlinear conservative potential can be scaled out by the renormalization of the amplitude of the solution. Therefore in what follows this value is chosen to be fixed.

We start the analysis of the particular cases with the natural test example where  $A(x)$ is chosen to be a solution of the standard NLS equation, i.e. having constant real linear and nonlinear potentials. We respectively set $V_l =$const  and $V_{nl}=0$. By means of straightforward algebra one obtains,  from (\ref{auxil_1}) and (\ref{w}), the relation $W_l(x) = -W_{nl}(x)A^2$. As it is clear this relation includes the particular case  $W_l = W_{nl}=0$, which precisely corresponds to the integrable NLS equation. Note, however, that the obtained relation allows for a general class of equations, all having the chosen function $A(x)$ as a solution  and  the ratio of the linear and nonlinear complex parts of the potential fixed to $-A^2(x)$.

Now we turn to the case where $U_l(x)\neq 0$ and $U_{nl}(x)$ is a real constant, which will be chosen to be $\pm 1$, i.e. to the case where the inhomogeneity is only linear and nonlinear gain or loss are absent.
The first example is a cnoidal wave (solution of {\it Type 3} according to the classification introduced in the previous section)
\begin{eqnarray}
\label{cn_focus}
A = A_0 \mbox{cn}(x,k),
\end{eqnarray}
embedded in the  linear lattice potential $V_l(x)= V_0 \mbox{cn}^2(x, k)$ with a constant focusing nonlinearity $\sigma = -1$. Requiring $\omega = 1/2-k^2$,  the respective pattern can be created with only the help of the linear dissipative term
\begin{eqnarray}
\nonumber
W_l=- W_0\mbox{sn}(x,k)\mbox{dn}(x,k),
\quad  W_0= \frac{3}{\sqrt{2}} \sqrt{A_0^2-V_0-k^2},
\label{case_1a}
\end{eqnarray}
as it follows from (\ref{w}).
The hydrodynamic velocity of the obtained solution is given by $v=\sqrt{2}W_0\mbox{cn}(x, k)$. The phase is: $\theta = (\sqrt{2}W_0/k)\mbox{arccos}(\mbox{dn}(x, x))$.
 The obtained  complex potential $U_l(x)$ is ${\cal PT}$-invariant.

\begin{figure}[ptb]
\setlength{\epsfxsize}{3in} \centerline{\epsfbox{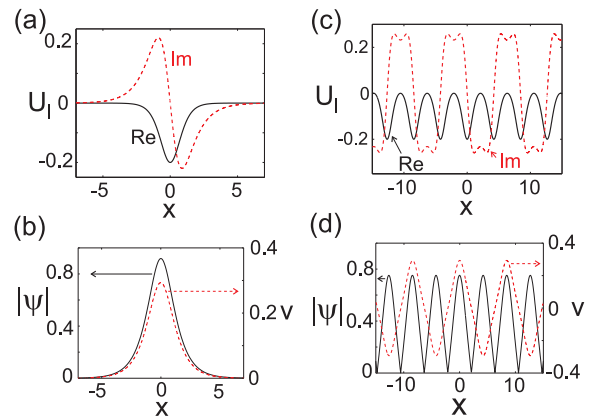}}\caption{ (color
online) (a) The distributions of the real (solid black) and imaginary (dashed red) parts of the linear potential given by Eq.(\ref{case_1a})
for $k=1$ and $A_0=0.92$, (b) the distribution of the amplitude $|\psi|$ (solid black, left vertical axis) and the phase gradient $v$ (dashed red, right vertical axis) of field $\psi$.
(c),(d) show the same but for $k=0.85$ and $A_0=0.75$. All panels are for $V_0=-0.2$ and $W_0=0.44$.}%
\label{figure1}%
\end{figure}

As it follows from (\ref{case_1a}) the nonlinear pattern $A(x)$ exists only for the amplitudes above the threshold $A_0^{(th)}=\sqrt{V_0+k^2}$ and grows with the intensity of the dissipative part.
If $k \neq 0$ then the solution is periodic. However if $k=1$ then our
solution is localized and has no zeros. So if $k=1$ then the solution belongs to { \it Type 2}. The typical field
distributions for solutions  are shown in Fig.~\ref{figure1}.

Let's now consider the stability of the solutions. To do this we linearize Eq.~(\ref{sys}) around the examined soliton solution.  The solution of the linearized equation can be sought in the form $\varphi(t, x)=\sum_m \varphi_{m}(x)exp(\lambda_m t)$. The spectrum $\lambda$ of the small perturbations has both  a continuous part, associated to non-localized eigenfunction and a discrete part associated to localized modes $\varphi_m$.
The existence of $\lambda$ with positive real part means that the corresponding eigenmode will grow exponentially in time and, thus, the examined solution will be unstable. In this paper we study the stability numerically substituting the spatial derivatives by their discrete analogs (we used 5-point approximation). Then the problem of finding the eigenvalues $\lambda$ governing the stability of the solution is reduced to the diagonlization of a matrix.

\begin{figure}[ptb]
\setlength{\epsfxsize}{3in} \centerline{\epsfbox{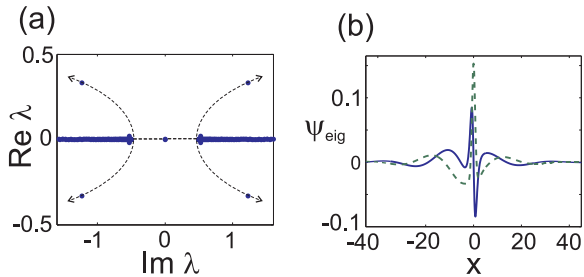}}\caption{ (color
online) (a) The spectrum for $k=1$, $A_0=1.04$, $V_{0}=-1$ and $W_{0}=2.2$ for the case when potentials are given by (\ref{case_1a}) and the background solution is given by Eq.(\ref{cn_focus}). The arrows show the direction of the motion of the eigenvalues when $V_{0,l}$ and
$W_{0,l}$ increase (b) The eigenvector of the unstable mode.}%
\label{figure2}%
\end{figure}

Discussing the stability we should first notice that if $V_{0}=W_{0}=0$ then Eq. (\ref{sys}) is nothing but the standard NLS equation. The spectrum of small perturbation on the background of the Schr\"odinger soliton has two pairs of degenerate zero eigenvalues corresponding to phase and translational symmetries. The introduction of a nonzero $U_l(x)$  preserves the phase invariance but breaks the translational symmetry. In Fig.~\ref{figure2} we show the evolution of the eigenvalues in the complex plane as the  linear potentials are increased with the ratio $V_{0}/W_{0}$ kept constant.

We see that by increasing the potentials  away from zero, the splitting of the zero eigenvalue present at $V_{0}=W_{0}=0$  gives a pair  of eigenvalues lying in the gap of the continuum which move in opposite directions along the imaginary axis without producing any instability of the solution. We have therefore that for sufficiently small $U(x)$ the soliton solution is always stable. However, when the pair of the eigenvalues reach the border and collide with the continuum they generate a quartet of complex eigenvalues with two unstable modes generating an oscillatory instability of the soliton.

\begin{figure}[ptb]
\setlength{\epsfxsize}{3in} \centerline{\epsfbox{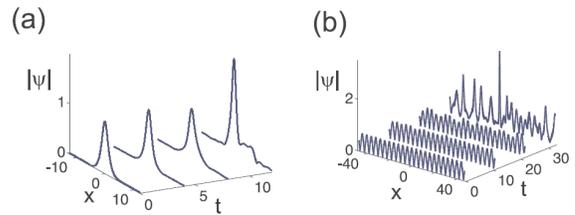}}\caption{ (color
online) (a) The decay of an unstable solitary solution (\ref{cn_focus}), $A_0=1.04$, $V_{0}=-1$ $W_{0}=2.2$, $k=1$ (b) The decay of an unstable periodical solution,
$A_0=0.7$, $V_{0}=-0.2$, $W_{0}=0.44$, $k=0.8$.}%
\label{figure3}%
\end{figure}

The development of the instability is illustrated in Fig.~\ref{figure3}. Panel (a) shows the instability of a soliton with the parameters $V_{0}=-1$ and $W_{0}=2.2$. The instability results in an infinitely growing localized peak in the area where the gain is positive. The formation of this peak is clearly seen on the picture. Panel (b) shows the decay of a periodic structure because of the instability present at the parameters $V_{0}=-0.2$, $W_{0}=0.44$ (let us mention that at this parameters localized solution is stable). One can see that the instability is of the same kind and results in the formation of growing peaks in the areas of positive gain.

\begin{figure}[ptb]
\setlength{\epsfxsize}{3in} \centerline{\epsfbox{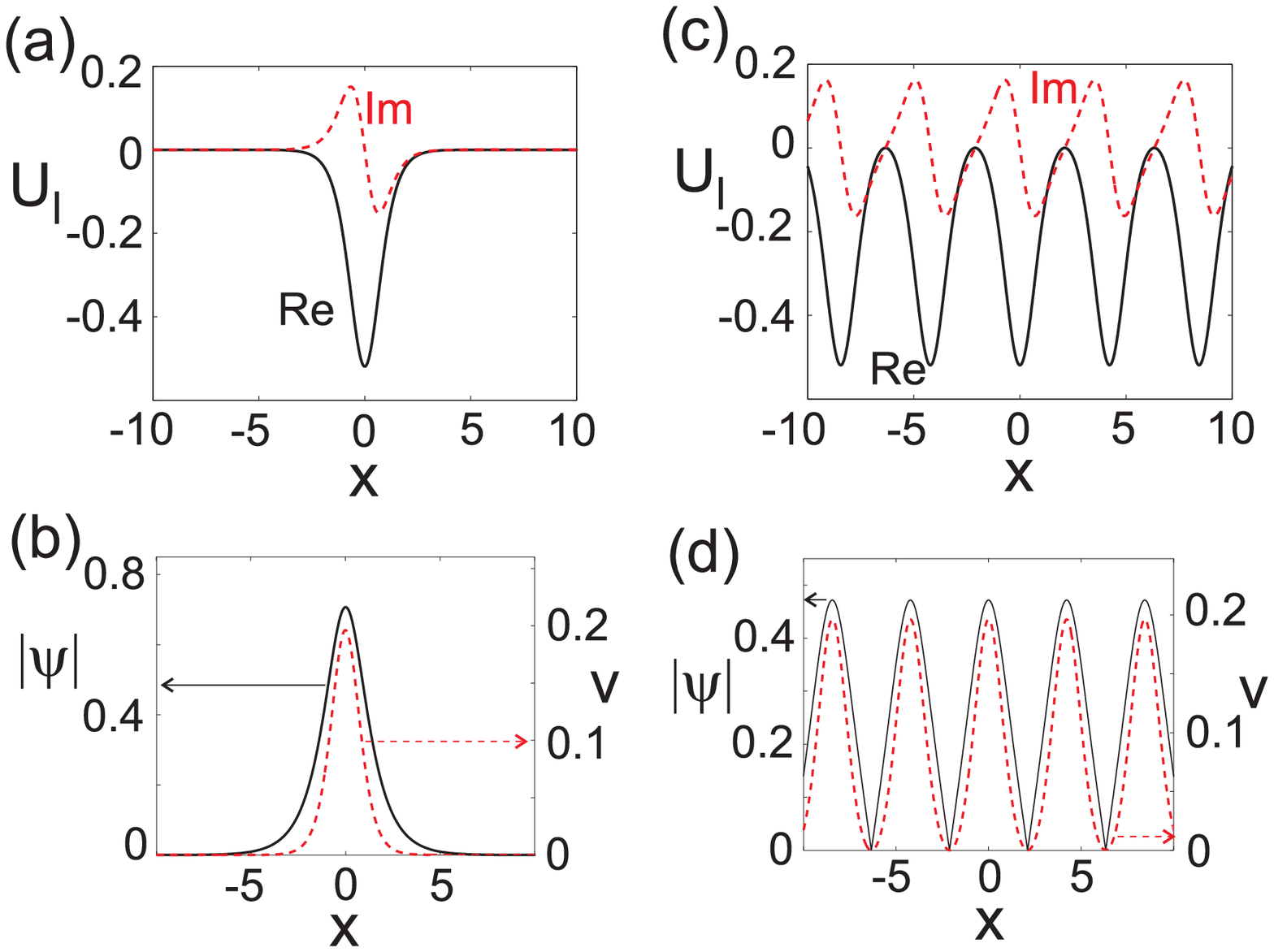}}
\caption{ (color
online) (a) The distributions of the real (solid black) and imaginary (dashed red) parts of the linear potential given by Eq.(\ref{eq11}) for $k=1$, $A_0=0.71$,
(b) the distribution of the amplitude of $|\psi|$ (solid black, left vertical axis) and the phase gradient (dashed red, right vertical axis) of field $\psi$ given by Eq.(\ref{cn_focus}). (c),(d) show the same but for
$k=0.85$, $A_0=0.47$. All panels are for $V_{0}=-0.5$ and $\tilde{V}=0.019$.}%
\label{figure4}%
\end{figure}

The solution (\ref{cn_focus}) is periodic and describes a pattern with spatially alternating currents, such that the average hydrodynamic velocity is zero (see Fig.~\ref{figure1}). It is not difficult to present an example where a pattern corresponds to a nonzero currents. To this end we consider the same wave (\ref{cn_focus}) but now in the potential
\begin{equation}
\label{eq11}
V_l(x)= V_0\mbox{cn}^2(x,k) - \tilde{V}\mbox{cn}^4(x,k),
\end{equation}
and require the wave amplitude to be $A_0=\sqrt{V_0+k^2}$ and frequency to be $\omega=1/2-k^2$. This immediately leads to the dissipative part of the potential in the form
$$W_l(x)= -\sqrt{8 \tilde{V}} \mbox{sn}(x)\mbox{cn}(x)\mbox{dn}(x). $$
Now the hydrodynamic velocity is given by $v(x)=\frac{W_0}{2} \mbox{cn}^2(x, k)$ and the phase is :
$$\theta = (W_0/2)E[\mbox{am}(x,k),k],$$
where $E(x, k)$ is the incomplete elliptic integral of the second kind.
   The obtained solutions can be either stable or unstable, the potential and field distributions for a stable solutions are shown in Fig.\ref{figure4}.

\begin{figure}[ptb]
\setlength{\epsfxsize}{3in} \centerline{\epsfbox{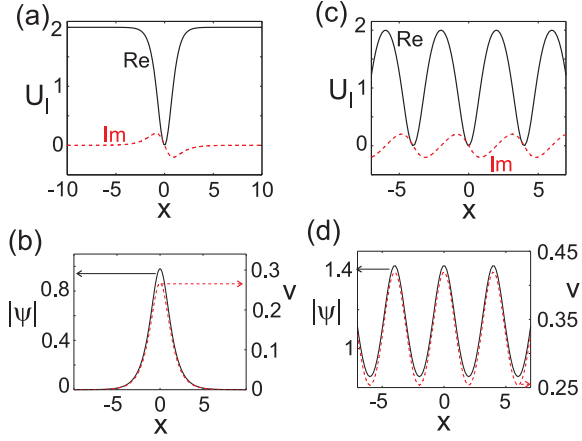}}\caption{ (color
online) (a) The distributions of the real (solid black) and imaginary (dashed red) parts of the linear potential given by Eq.(\ref{dn_pt})
for $k=1$, $A_0=0.98$, (b) the distribution of the amplitude $|\psi |$ (solid black, left vertical axis) and the phase gradient $v$ (dashed red, right vertical axis) of the field $\psi$ given by (\ref{dn}). (c),(d)
show the same but for $k=0.8$, $A_0=1.43$. All panels are
for $V_{0,l}=2$ and $W_{0,l}=0.4042$. }%
\label{figure5}%
\end{figure}

Let us now turn to the case of defocusing medium: $\sigma \equiv 1$ and concentrate on the conservative potential in the form $V_l=V_0\mbox{sn}^2(x,k)$. A stable wave can be chosen in the form
\begin{eqnarray}
\label{dn}
A=A_0 \mbox{dn(x,k)}
\end{eqnarray}
and is induced by the spatial distributions of the dissipation and losses
\begin{eqnarray}
\nonumber W_l = -W_{0} \mbox{sn}(x,k)\mbox{cn}(x,k),\\
\label{dn_pt} W_0^2=\frac{9k^2}{2}\left(V_0-k^2-k^2A_0^2\right).
\end{eqnarray}
One can see that is solution is of {\it Type 1}. The frequency of the solutions and the hydrodynamic velocity are given by $\omega=V_0/k^2-1+k^2/2$ and $v(x)=\frac{2W_0}{3k^2}\mbox{dn}(x)$.
The phase then is readily obtained from the velocity as $\theta(x)=\int_{-\infty}^x  v(y)dy$.
Fig.\ref{figure5} illustrates the potential and the field distribution for a stable solution of such a kind.

\section{Case of nonlinear complex potentials}

\begin{figure}[ptb]
\setlength{\epsfxsize}{3in} \centerline{\epsfbox{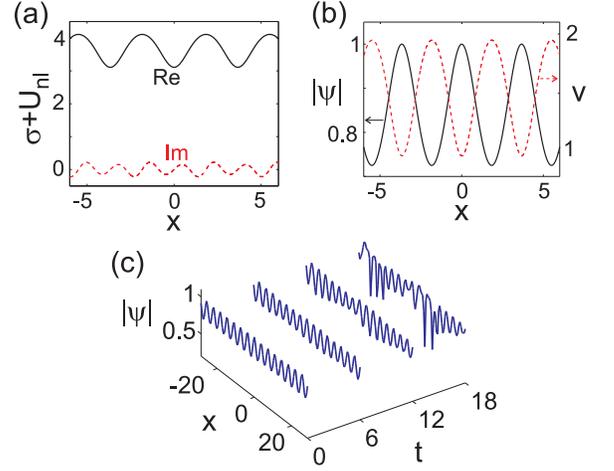}}\caption{ (color
online) (a) The distributions of the real (solid black) and imaginary (dashed red) parts of the nonlinear potential given by Eq.(\ref{eq20}),(\ref{eq20a}),
(b) the distributions of the amplitude $|\psi|$ (solid black, left vertical axis) and the phase gradient (dashed red, right vertical axis) of the field $\psi$.
(c) The instability developing on the solution (b) perturbed with small noise. All panels are for $k=0.689$.}%
\label{figure6}%
\end{figure}

Let us now turn to the case where the linear potential is absent, i.e. $U_l(x)\equiv 0$, and consider spatial dependent  nonlinearity. Respectively it is considered that $V_{nl}(x)$ is given and the  problem to design the dissipative term inducing the given field pattern is posed. Notice that in general  the conservative nonlinear part cannot change sign, i.e. the medium cannot be focusing in some regions of the domain and defocusing in others.  We consider the case of a defocusing nonlinearity $\sigma=1$ and look for a solution of the type  (\ref{dn}) with $A_0=1$ (a solution belonging to {\it Type 1}) embedded in the potential
\begin{eqnarray}
\label{eq20}
V_{nl}=\frac{1}{k^2} + \mbox{sn}^2(x,k),
\end{eqnarray}
then the hydrodynamic velocity is given by $v(x)=\sqrt{2}k \left( 1+ \mbox{sn}^2(x, k)\right)$ and imposing the frequency $\omega = 3k^2/2 + 1+1/k^2$
nonlinear dissipative part is computed from (\ref{w}) to be
\begin{eqnarray}
\label{eq20a}
W_{nl}=\frac{\sqrt{2}k \mbox{sn}(x,k)\mbox{cn}(x,k)}{\mbox{dn}^3(x,k)}(1-k^2-2k^2\mbox{sn}^2(x,k)).
\end{eqnarray}
The profiles of the nonlinear potential, the distribution of the field and the development of the instability for the solution (\ref{eq21}) are shown in Fig.\ref{figure6}.

In the case of {\it focusing} nonlinearity $\sigma \equiv -1$ we consider the mode (\ref{cn_focus}) (belonging to {\it Type 3}) loaded in the potential
\begin{eqnarray}
V_{nl}(x)= 1-\frac{1 + k^2}{2A_0^2} + \tilde{V}_{nl} \mbox{cn}^2(x,k)
\end{eqnarray}
and set $\omega = 1/2-k^2/2$. Then, from (\ref{w}) we get
\begin{equation}
W_{nl}(x) = -\frac{\sqrt{8\tilde{V}_{nl}}}{A_0}\frac{\mbox{sn}(x)\mbox{dn}(x)}{\mbox{cn}(x)}.
\end{equation}
The phase is: $\theta = \sqrt{2\tilde{V_{nl}}}A_0 E[\mbox{am}(x,k),k]$.

\begin{figure}[ptb]
\setlength{\epsfxsize}{3in} \centerline{\epsfbox{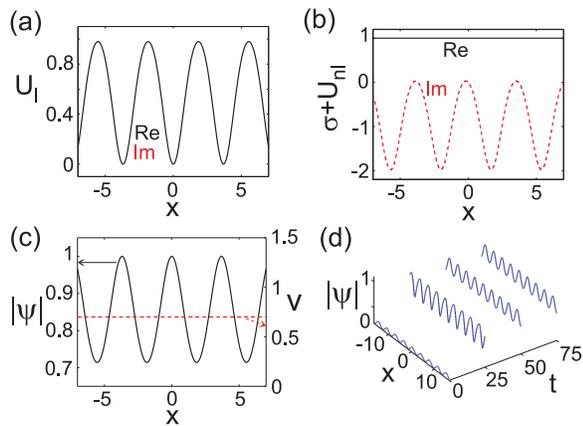}}\caption{ (color
online) (a) The distributions of the real and imaginary parts of the linear potential given by Eq. (\ref{eq21}) for $k=0.7$ (they coincide),
(b) the same for the nonlinear potential given by (\ref{eq21a}), solid black line is for the real pard and dashed red line is for the imaginary part, (c) the distribution of the amplitude $|\psi|$ (solid black, left vertical axis) and the phase gradient $v$ (dashed red, right vertical axis) of the
field $\psi$. (d) The relaxation of the initial field to the solution shown in (c). All panels are for $a=1$ and $k=0.7$. }%
\label{figure7}%
\end{figure}

Finally we consider a more general case when there exists a linear real potential with complex linear and nonlinear inhomogeneous dissipative terms. We let $\sigma \equiv 1$ and consider the pattern (\ref{dn}) loaded in the linear lattice $V_l(x)=2k^2\mbox{sn}^2(x,k)$. Then  requiring   $\omega =  k^2 +1$ we obtain that the pattern (\ref{dn})  of the {\it Type 1} is induced by the  distribution of linear and nonlinear dissipative terms as follows
\begin{eqnarray}\label{eq21}
W_l(x)\equiv a V_l(x),
\\
\label{eq21a}
W_{nl} = -\frac{k^2 \mbox{sn}(x,k)}{ \mbox{dn}^3(x)}\left[2a \mbox{sn}(x,k)\mbox{dn}(x,k) +k  \mbox{cn}(x,k)\right].
\end{eqnarray}
 The phase is $\theta = k x$ and the velocity field is therefore constant.
The potentials and the field distributions for Eqs.(\ref{eq21}),(\ref{eq21a}) when $a=1$ are shown in Fig.~\ref{figure7}.
The solution
can be stable in its existence  $V_{0l}>k^2$ with $V_{0l}$ the  amplitude of the real linear  potential.
The evolution is shown in panel (d) of Fig.~\ref{figure7}.

\section{Dissipative breathers}

\begin{figure}
\begin{center}
    \includegraphics[width=3.5cm]{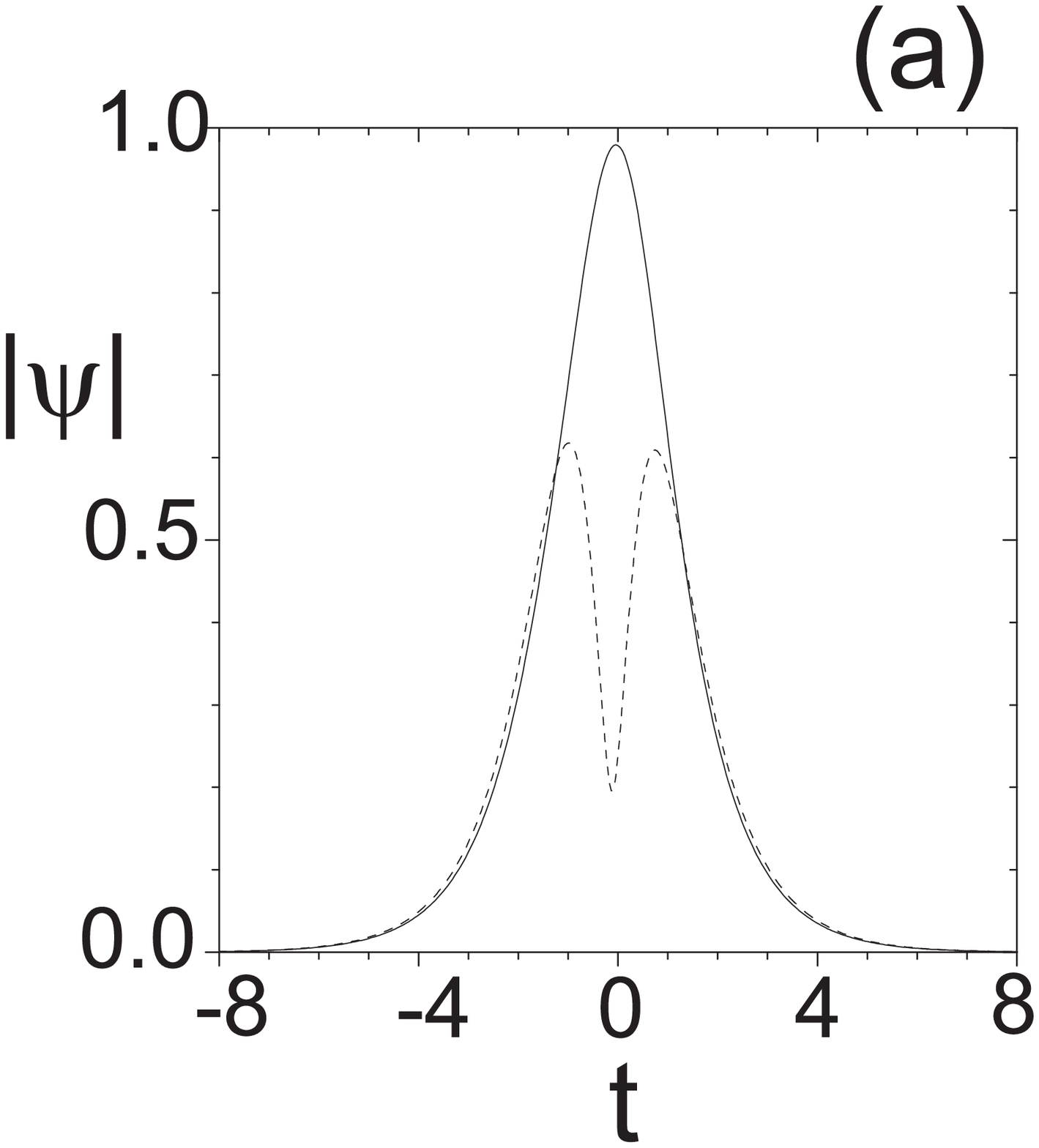}
    \includegraphics[width=3.5cm]{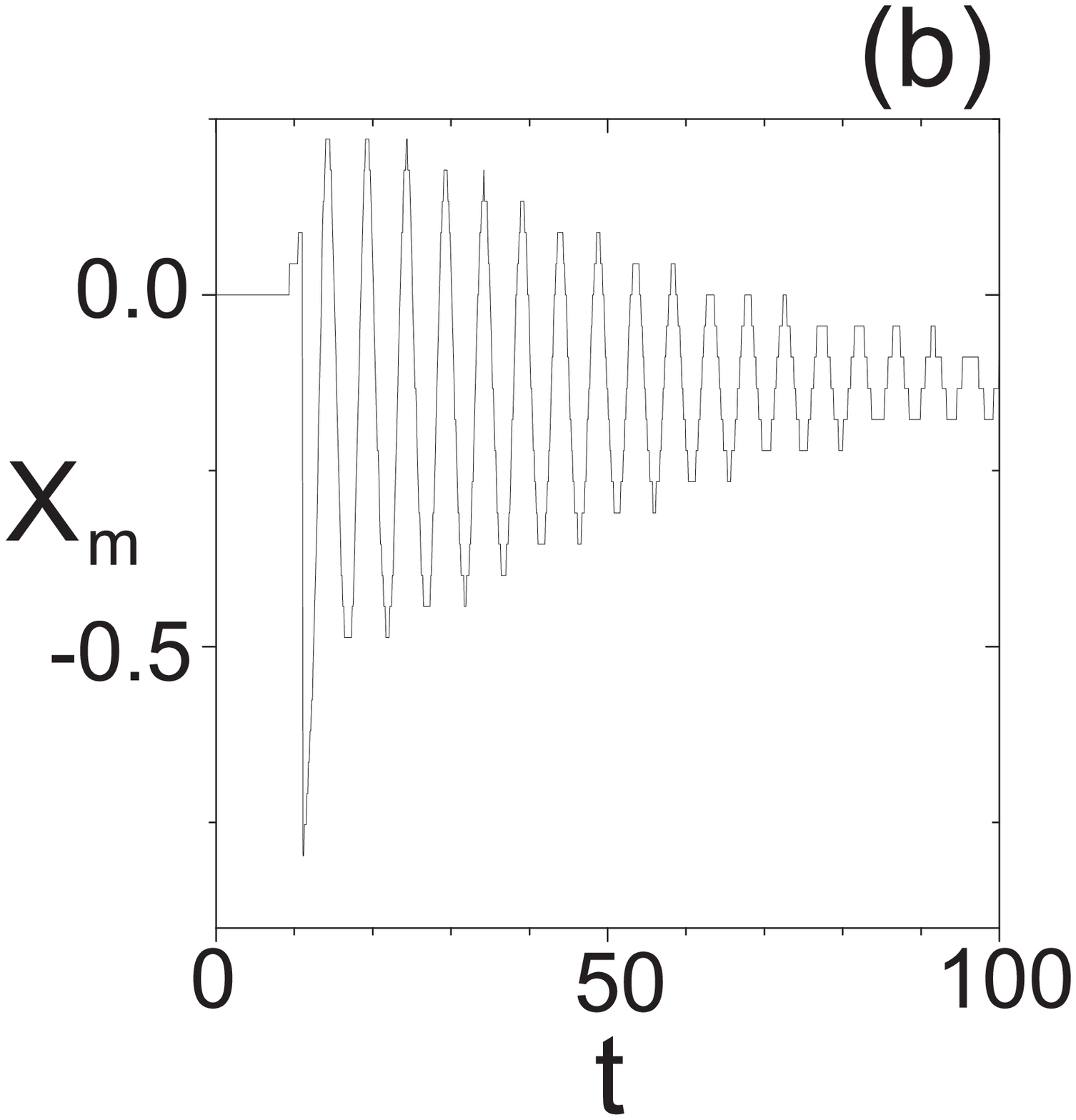}
\end{center}
\begin{center}
    \includegraphics[width=3.5cm]{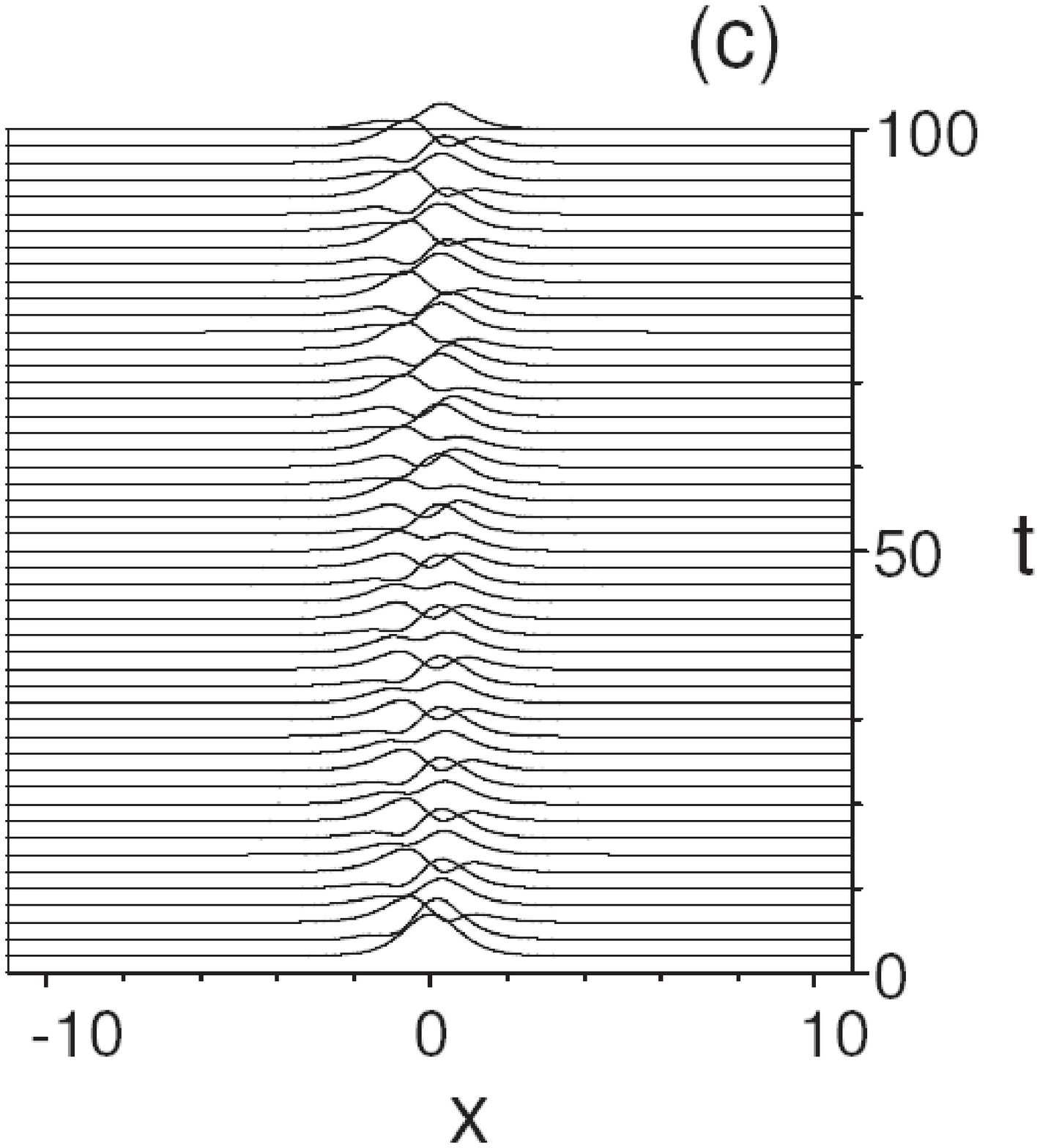}
    \includegraphics[width=3.5cm]{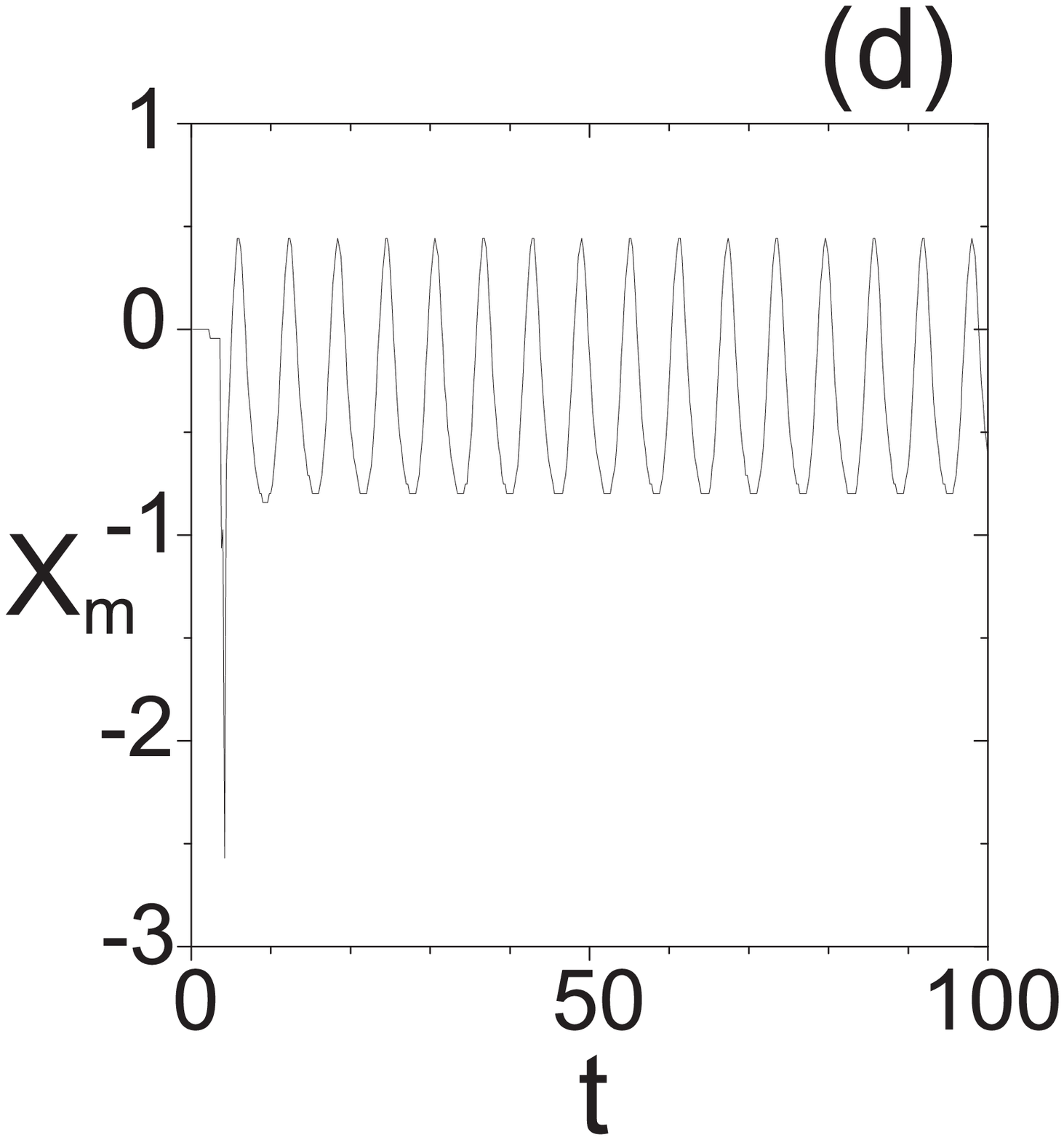}
\end{center}
\caption{ (a) The modulus of the wavefunction before ($\Omega=0.05$, continuous line)  and after ($\Omega=0.075$, dashed  line) the bifurcation.
(b) Relaxation of the position of the relative minimum of the density to the new stationary equilibrium, (c) Time evolution of the breating mode at $\Omega=0.25$. (d) The positition of the relative minimum of the density profile versus time during the breather oscillations. All panels are for the same parameters values as in Fig. 7 except for $a=2.0$ and $k=1$.}%
\label{figure8}%
\end{figure}

In the  limit of infinite period (e.g. for modulus $k=1$) we have that the potentials in Eqs. (\ref{eq21})-(\ref{eq21a}) give rise  to a stable localized hump in the sense that any perturbation of it will relax back to the stationary state. Stable solutions of this type correspond to attractive centers. It is possible to destabilize it into limit cycles and create stable dissipative breathers, by analogy with the bifurcation of a dissipative soliton to a pulsating soliton in the complex Ginzburg-Landau equation  ~\cite{AST} (similar studies can be done with other types of soliton solutions given above). We take this solution  as initial condition of the complex NLS  with a parabolic trap potential added to the real linear potential, e.g. we take $V_l(x)=\frac 12 \Omega^2 x^2 + 2k^2\mbox{sn}^2(x,k)$, keeping all other potentials the same. We use the frequency of the parabolic trap as a parameter to induce the transition into a  limit cycle. For  $\Omega$ small enough the extra potential will act as a small perturbation allowing  the solution to relax to a nearby stationary solution of similar shape. We find, however, that when $\Omega$ exceeds a critical value the solution bifurcates from a single hump to a double humps shape which is stable and stationary e.g. it still corresponds to an attractive center. In the panel (a) of Fig.~\ref{figure8} we show the shapes of the solutions taken before and after the bifurcation, while in panel (b) we depict the relaxation dynamics of the center of the two humps solution toward the new equilibrium  center. As we increase the strength of the parabolic trap the relaxation time increases until it becomes a regular periodic oscillation (see panel (d)). The breather exists in a window in trap frequencies that, for the given parameters, starts at $\Omega_1\approx 0.2075$ and  ends at $\Omega_2 \approx 0.27174$. Outside this window the solution returns to be stationary.    In the panel (c) of Fig.~\ref{figure8} we have shown the time evolution of the density profile during the  breather motion for a parameter value inside the above existence  window.

\section{Conclusions}

Summarizing, we have found the exact solitonic and nonlinear periodic solutions for the generalized NLS equation
with complex linear and nonlinear potentials. We have confirmed numerically the stability of some solutions, in particular those having the PT symmetry.
We have shown that if stable, the solutions are quite robust against perturbations and in this sense  they are generic. We have  investigated stable
breather solutions which originate when a parabolic trap is switched on as additional real potential. Using the strength of the parabolic trap as parameter we
have demonstrated that the stationary solutions undergo a bifurcation to a limit cycle which correspond to a stable breather mode.
The stability of both stationary periodic and single hump soliton solutions and breathers   opens the possibility to
experimentally observe such waves both in nonlinear optics and in BEC with optical lattices in presence of dissipative effects.

\bigskip

\section*{Acknowledgements} FKA and VVK were  supported by  the 7th European Community Framework Programme
under the grant PIIF-GA-2009-236099 (NOMATOS). AVY was partially supported by the FCT grant PTDC/EEA-TEL/105254/2008.
MS acknowledges partial support from MIUR through  a PRIN-2008
initiative. This cooperative work is also partially  supported by a bilateral project 2009-2010 within the framework of the
Portugal (FCT) - Italy (CNR) agreement. The authors are grateful to R.M. Galimzyanov for useful discussions.

\end{document}